\documentclass[preprint,preprintnumbers,amsmath,showpacs,amssymb]{revtex4}

\usepackage{epsfig,graphicx}
\usepackage{dcolumn}
\usepackage{bm}
\begin{document}
\draft

\title{Properties of weighted complex networks}
\author{Xin-Jian Xu, Zhi-Xi Wu, and Ying-Hai Wang\footnote{For correspondence: yhwang@lzu.edu.cn}}
\address{Institute of Theoretical Physics, Lanzhou University, Lanzhou Gansu 730000, China}

\date\today

\begin{abstract}
We study two kinds of weighted networks, weighted small-world
(WSW) and weighted scale-free (WSF). The weight $w_{ij}$ of a link
between nodes $i$ and $j$ in the network is defined as the product
of endpoint node degrees; that is $w_{ij}=(k_{i}k_{j})^{\theta}$.
In contrast to adding weights to links during networks being
constructed, we only consider weights depending on the \lq\lq
popularity\rq\rq of the nodes represented by their connectivity.
It was found that the both weighted networks have broad
distributions on characterization the link weight, vertex
strength, and average shortest path length. Furthermore, as a
survey of the model, the epidemic spreading process in both
weighted networks was studied based on the standard
\emph{susceptible-infected} (SI) model. The spreading velocity
reaches a peak very quickly after the infection outbreaks and an
exponential decay was found in the long time propagation.
\end{abstract}


\maketitle

\section{Introduction}
Complex networks have attracted an increasing interest recently.
The main reason is that they play an important role in the
understanding of complex behaviors in real world networks \cite
{Strogatz, Barabasi_1, Dorogovtsev, Newman_0}, including the
structure of language \cite {Cancho, Sigman}, scientific
collaboration networks \cite {Newman_1, Barabasi_2}, the Internet
\cite {Albert, Huberman_1} and World Wide Web \cite {Huberman_2,
Caldarelli}, power grids \cite {Amaral}, food webs \cite {McCann,
Williams}, chemical reaction networks \cite {Alon}, metabolic
\cite {Jeong_1} and protein networks \cite {Jeong_2}, etc. The
highly heterogeneous topology of these networks is mainly
reflected in two characters, the small average path lengths among
any two nodes (small-world property) \cite {Watts_1} and a power
law distribution (scale-free property), $P(k)\sim k^{- \gamma}$
with $2 \leq \gamma \leq 3$, for the probability that any node has
$k$ connections to other nodes \cite {Barabasi_3}.

Most of these networks that have been studied in the physics
literature have been binary in nature; that is, the edges between
vertices are either present or not.  Such networks can be
represented by $(0,1)$ or binary matrices.  A network with $N$
vertices is represented by an $N \times N$ adjacency matrix
\textbf{W} with elements
\begin{equation}
w_{ij}=\biggl\lbrace
\begin{array}{ll}
1 & \quad \mbox{if $i$ and $j$ are connected,}\\
0 & \quad \mbox{otherwise.}
\end{array}\tag{1}\label{eq1}
\end{equation}
However, many networks are intrinsically weighted, their edges
having differing strengths. In a social network there may be
stronger or weaker social ties between individuals.  In a
metabolic network there may be more or less flux along particular
reaction pathways.  In a food web there may be more or less energy
or carbon flow between predator-prey pairs.  Edge weights in
networks have received relatively little attention in the physics
literature for the excellent reason that in any field one is well
advised to look at the simple cases first (unweighted networks)
before moving on to more complex ones (weighted networks).

On the other hand, there are many cases where edge weights are
known for networks, and to ignore them is to throw out a lot of
data that, in theory at least, could help us to understand these
systems better. Recently some weighted network models are
proposed. In Ref. \cite {Yook}, a weighted network model was
presented, in which both the structure and the connection weights
are driven by the connectivity according to the preferential
attachment rule. In Ref. \cite {Zheng}, a weighted evolving
network model with stochastic weight assignments was suggested. In
Ref. \cite {Barrat_1}, it was also found that the strengths of
nodes in complex networks obey a power-law distribution. In Ref.
\cite{Park}, a weighted network was introduced to account for
possible distinct functional roles played by different nodes and
links in complex networks by utilizing the concept of betweenness.

It should be noted that, most of these existing weighted network
models are driven entirely by the preferential attachment scheme
in structure and the preferential strengthening scheme in
connection weights. In this paper, we will introduce a class of
weight to two prototype complex networks, the Watts-Strogatz (WS)
model and the Barab\'{a}si-Albert (BA) model, considering link
weights depending on the \lq\lq popularity\rq\rq of the nodes
represented by their connectivity. We shall show that the two
weighted networks, weighted WS (WWS) and weighted BA (WBA), have
broad distributions on characterization their structures and new
results of the spread of infectious diseases.

\section{The model}
In real systems, a link's weight reflect the familiarity between
two nodes, which can be regarded as the combination of the nodes'
\lq\lq popularity\rq\rq. For example, in the airport network, each
given weight $w_{ij}$ is the number of available seats on direct
flight connections between the airports $i$ and $j$; in the
scientific collaboration network, the nodes are identified with
authors and the weight depends on the number of coauthored papers.
Based on this simple mechanism, we will assign weights to links in
two prototype complex networks, the WS and BA graphs. The two
models, which reveal the small-world and scale-free properties
respectively, have become the most popular models to be used to
study the statistics of many real networks and explore the
dynamics on the network structures.

\emph{WS graph}: Starting with a ring of $N$ vertices, each
connected to its $2K$ nearest neighbors by undirected edges, and
then each local link is visited once with the rewiring probability
$p$ it is removed and reconnected to a randomly chosen node.
Duplicate and self-connected edges are forbidden. After the whole
sweep of the entire network, a small-world graph is constructed
with an average connectivity $\langle k \rangle = 2K$ (in the
present work we will consider the parameters $p=0.1$ and $K=5$).

\emph{BA graph}: Starting from a small number $m_{0}$ of nodes,
every time step a new vertex is added, with $m$ links that are
connected to an old node $i$ with probability
\begin{equation}
\Pi_{i}=\frac{k_{i}}{\sum_{j}k_{j}},\tag{2}\label{eq2}
\end{equation}
where $k_{i}$ is the connectivity of the $i$th node. After
iterating this scheme a sufficient number of times, we obtain a
network composed by $N$ nodes with connectivity distribution $P(k)
\sim k^{-3}$ and average connectivity $\langle k \rangle =2m$ (in
the present work we will consider the parameters $m_{0}=m=3$).

After the networks have been constructed, we shall define the
weight $w$ of a link between nodes $i$ and $j$ in the network to
be the product of endpoint node degrees
\begin{equation}
w_{ij}=\biggl\lbrace
\begin{array}{ll}
(k_{i}k_{j})^{\theta} & \quad \mbox{if $i$ and $j$ are connected,}\\
0 & \quad \mbox{otherwise}
\end{array}\tag{3}\label{eq3}
\end{equation}
where $\theta$ is a positive constant. In contrast to adding
weights to links during networks being constructed, we only
consider weights depending on the nodes connectivity.

\section{Structural properties}
\begin{figure}[h]
\centerline{\epsfxsize=15cm \epsffile{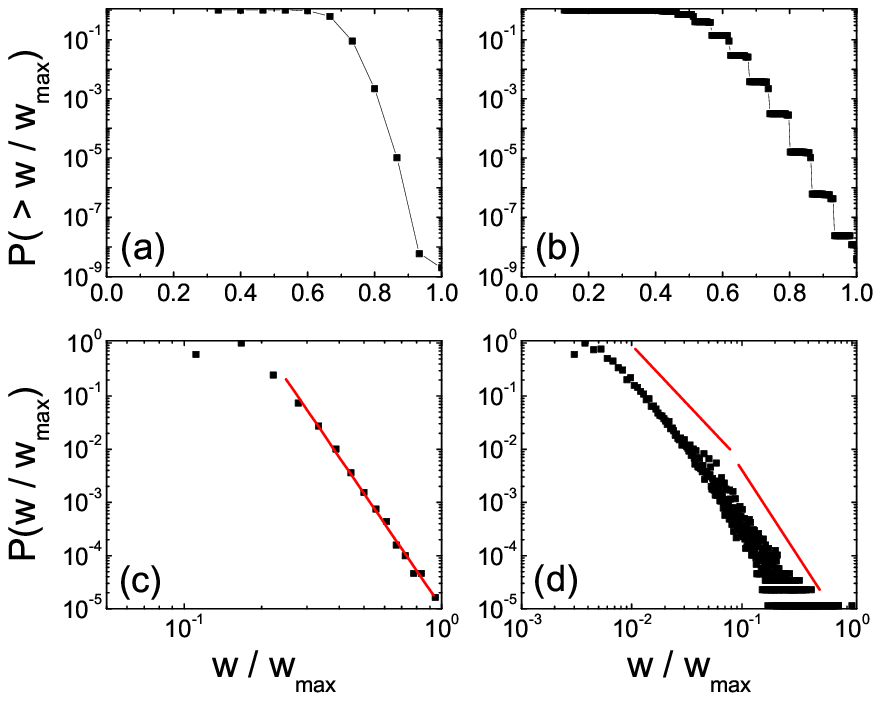}}

\caption{Linear-log plot of the cumulative distribution
$P(x>w/w_{max})$ of the scaling individual link weight in the WWS
network with $\theta=0.5$ (a) and $\theta=1$ (b) and the log-log
plot of the distribution $P(w/w_{max})$ of the scaling individual
link weight in the WBA network with $\theta=0.2$ (c) and
$\theta=0.5$ (d). The data were averaged on 10 networks of size
$N=10^{6}$.} \label{fig1}
\end{figure}

The standard topological characterization of networks is obtained
by the analysis of the probability distribution $P(k)$ that a
vertex has degree $k$. Similarly, the first topological
characterization of weights is obtained by the distribution
$P(w)$, which is defined as the probability that a randomly
selected link between two nodes has a weight value $w$. In Fig.
\ref{fig1}(a)-(b), we plot the cumulative distribution
$P(x>w/w_{max})$ on linear-log scale for WWS networks. The
cumulative distribution is defined as
\begin{equation}
P(>x)=\sum_{x^{\prime}=x}p(x^{\prime}).\tag{4}\label{eq4}
\end{equation}
The both distributions decay faster than an exponential decay
would. In Fig. \ref{fig1}(c)-(d), we plot the distribution
$P(w/w_{max})$ on log-log scale for WBA networks. Fig.
\ref{fig1}(c) shows a truncated power law behavior with the
apparent exponent $\alpha = -7.10 \pm 0.08$. Fig. \ref{fig1}(d)
shows a double power law behavior with the apparent exponent
$\alpha = -2.45 \pm 0.03$ (for small-$w$) and $\alpha = -3.08 \pm
0.06$ (for large-$w$).

\begin{figure}[h]
\centerline{\epsfxsize=15cm \epsffile{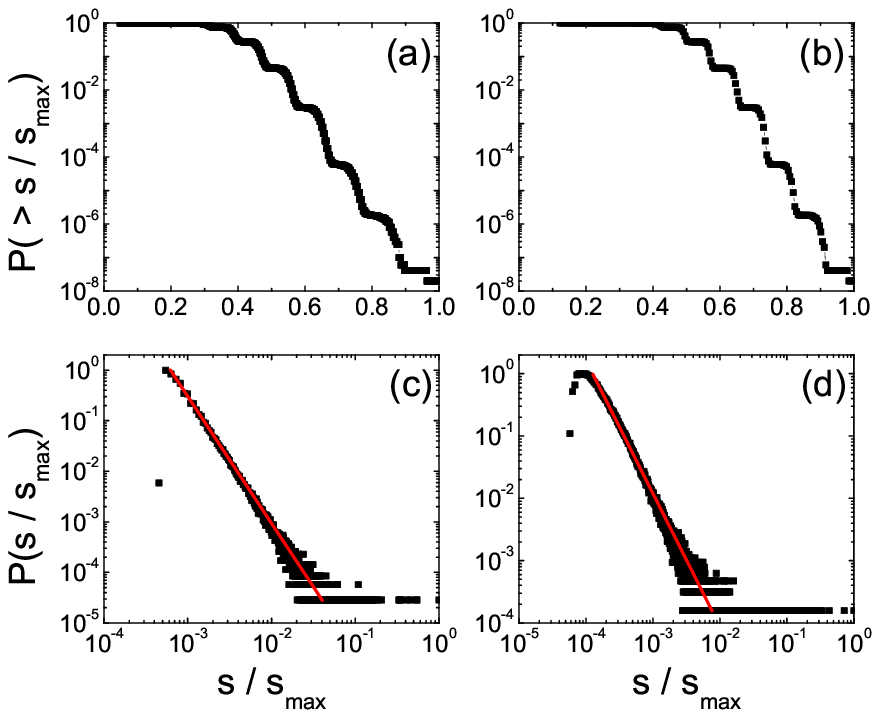}}

\caption{Linear-log plot of the cumulative distribution
$P(x>s/s_{max})$ of the scaling individual link weight in the
weighted WS network with $\theta=0.5$ (a) and $\theta=1$ (b) and
the log-log plot of the distribution $P(s/s_{max})$ of the scaling
individual link weight in the weighted BA network with
$\theta=0.2$ (c) and $\theta=0.5$ (d). The data were averaged on
10 networks of size $N=10^{6}$.} \label{fig2}
\end{figure}

Along with the degree of a node, a very significative measure of
the network in terms of the actual weights is obtained by looking
at the vertex strength $s_{i}$ defined as
\begin{equation}
s_{i}=\sum_{j \in U(i)}w_{ij}\tag{5}\label{eq5}
\end{equation}
where the sum runs over the set $U(i)$ of neighbors of $i$. The
strength of a node integrates the information both about its
connectivity and the importance of the weights of its links, and
can be considered as the natural generalization of the
connectivity. In Fig. \ref{fig2}(a)-(b), we plot the cumulative
distribution $P(x>s/s_{max})$ on linear-log scale for WWS
networks. The both distributions follow a roughly exponential
decay. In Fig. \ref{fig2}(c)-(d), we plot the distribution
$P(w/w_{max})$ on log-log scale for WBA networks. The both
distributions follow a power-law behavior with the apparent
exponent $\alpha = -2.66 \pm 0.02$ (Fig. \ref{fig2}(c)) and
$\alpha = -2.31 \pm 0.02$ (Fig. \ref{fig2}(d)).

\begin{figure}[h]
\centerline{\epsfxsize=15cm \epsffile{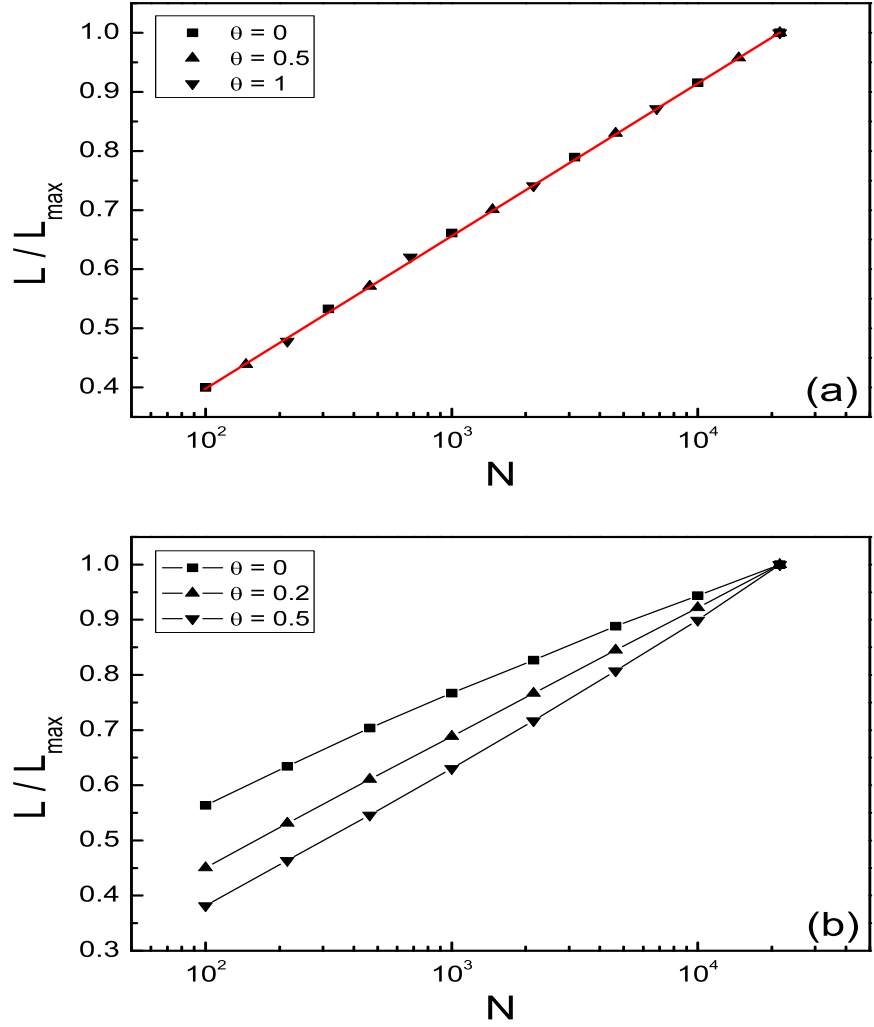}}

\caption{Scaling average shortest path length in the WWS (a) and
WBA (b) networks. Simulations were implemented on the networks
averaging over $50$ different realizations.} \label{fig3}
\end{figure}

The shortest path plays an important role for the transport within
a network \cite {Goh_1, Szabo}. A path denotes a sequence of
vertices, successive pairs of which are connected via edges. In
general there exist many paths connecting two given vertices. The
shortest path is the one with minimum path length among all the
paths. In weighted networks, the average shortest path length can
be defined as
\begin{equation}
L= \frac{2}{N(N-1)}\sum_{i,j}w_{i,j} \tag{6}\label{eq6}
\end{equation}
where $w_{i,j}$ is the the minimum value of the sum of weights
among all the paths between node $i$ and $j$. In Fig. \ref{fig3}
scaling average shortest path lengths of the weighted WS and
weighted BA networks are shown. The best linear fit in Fig.
\ref{fig3}(a) yields a slope $0.250 \pm 0.001$, which implies that
the behavior of the average shortest path length of the weighted
WS network remain invariable. Whereas, Fig. \ref{fig3}(b) yields
that the slope of the linear fit of the weighted BA network is
about $\langle k \rangle^{2\theta}$ times as that of the SF
network.

\section{Epidemic spreading}
In the study of complex networks, a good example is to inspect the
effect of their complex features on the dynamics of epidemic and
disease spreading. It is easy to foresee that the characterization
and understanding of epidemic dynamics on these networks can find
immediate applications to a large number of problems, such as
computer virus infections, distribution of wealth, transmission of
public opinion, etc. Recent papers \cite {Kuperman, Pastor,
Newman_3} have given some valuable insights of that: for
small-world networks, there is a critical threshold below which an
infection with a spreading rate dies out; on the contrary, for
scale-free networks, even an infection with a low spreading rate
will prevalence the entire population. However, so far, studies of
epidemic spreading just focus on unweighted networks, and a
detailed inspection of epidemic spreading in weighted networks is
missing.

To study the dynamics of infectious diseases spreading on weighted
networks, we shall study the standard \emph{susceptible-infected}
(SI) model \cite{Murray}. In this model individuals can only exist
in two different states: susceptible (or healthy) and infected.
The model can be described in terms of the densities of
susceptible and infected individuals, $s(t)$ and $i(t)$,
respectively, then $s(t)+i(t)=1$. Each individual is represented
by a vertex of the network and the links are the connections
between individuals along which the infection may spread. In
weighted networks, according to Ref. \cite {Yan}, the spreading
rate can be defined as
\begin{equation}
\lambda_{ij}=\frac{w_{ij}}{w_{max}}\tag{6}\label{eq6}
\end{equation}
at which susceptible individual $i$ acquire the infection from the
infected neighbor $j$, where $w_{max}$ is the largest value of
$w_{ij}$ in the network. In this model, infected individuals
remain always infective, an approximation that is useful to
describe early epidemic stages in which no control measures are
deployed.

\begin{figure}[h]
\centerline{\epsfxsize=15cm \epsffile{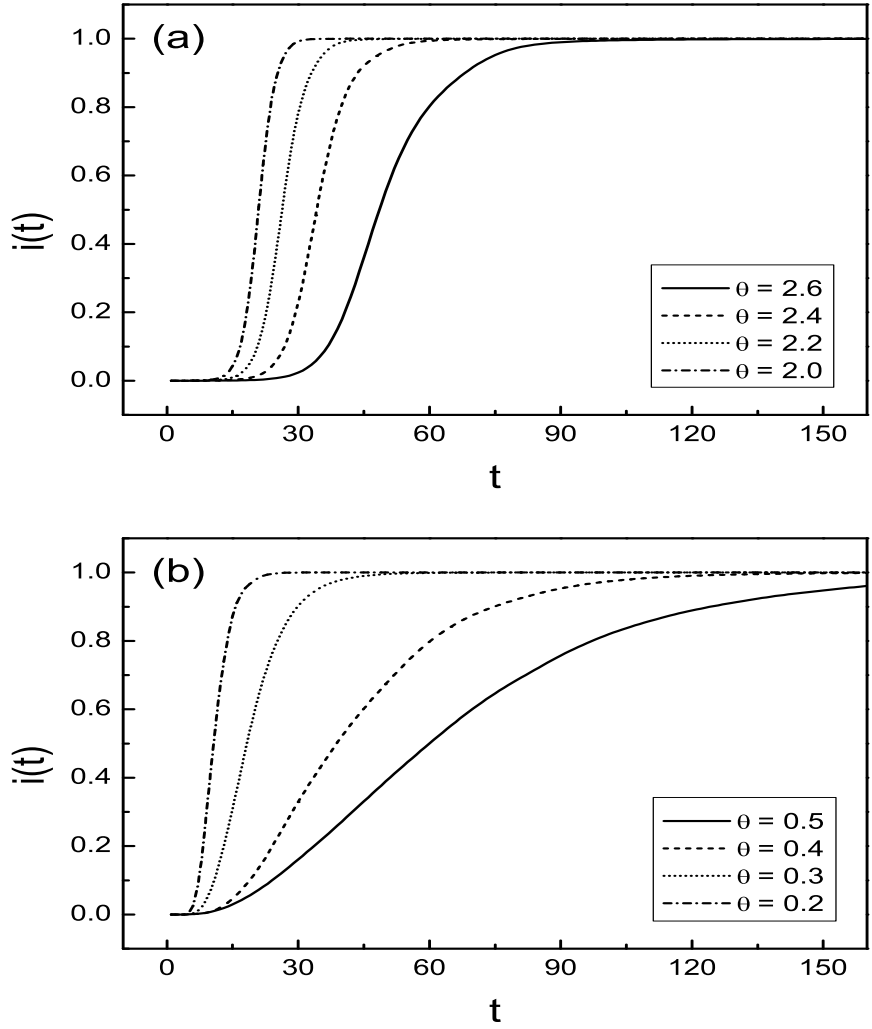}}

\caption{Density of infected individuals versus time in the WWS
(a) and WBA (b) networks with $N=10^{5}$. All the plots were
averaged over $100$ experiments.} \label{fig4}
\end{figure}

\begin{figure}[h]
\centerline{\epsfxsize=15cm \epsffile{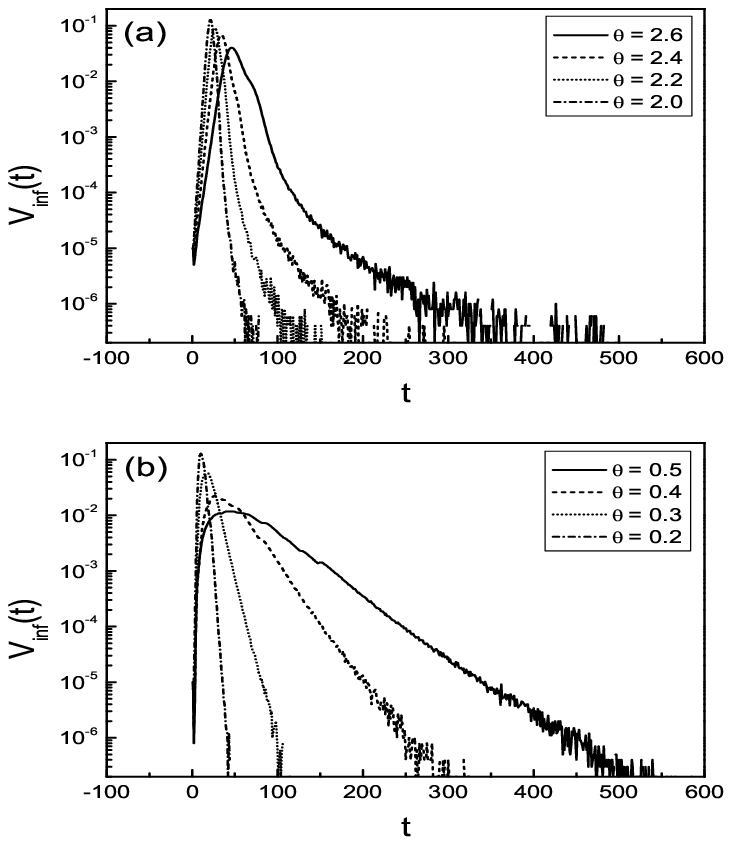}}

\caption{Spreading velocity at each time $t$ in the WWS (a) and
WBA (b) networks with $N=10^{5}$. All the plots were averaged over
$100$ experiments.} \label{fig5}
\end{figure}

We start simulations by selecting one vertex randomly and assuming
it is infected. The disease will spread in the network in
according with the rule of Eq. (\ref{eq6}). In Fig. \ref{fig4} we
plot the density of infected individuals versus time in both
weighted networks. Since $\frac{w_{ij}}{w_{max}} \leq 1$, the
smaller $\theta$ is, the more quickly the infection spreads. All
the individuals will be infected in the limit of long time as
$\lim_{t \rightarrow \infty}i(t)=1$. We will study the detail of
spreading velocity at the outbreak moment which is defined as
\begin{equation}
V_{inf}(t)=\frac{di(t)}{dt} \approx i(t)-i(t-1).\tag{7}\label{eq7}
\end{equation}
We account the number of newly infected vertices at each time step
and report the spreading velocity in Fig. \ref{fig5}. Apparently,
the spreading velocity goes up to a peak quickly and leave us very
short response time to develop control measures. At the moment of
infection outbreaks, the number of infected individuals is very
small, as well as a very long time from the outbreak, the number
of susceptible individuals is very small. Thus when $t$ is very
small (or large), the spreading velocity is close to zero.
Moreover, in contrast to trivial velocity decay for the WWS
network (see Fig. \ref{fig5}(a)), an exponential decay was found
for the WBA network (see Fig. \ref{fig5}(b)).

\section{Conclusions}
A more complete view of complex networks is provided by the study
of the interactions defining the links of the system. In real
systems, a link's weight reflect the familiarity between two
nodes, which can be regarded as the combination of the nodes'
\lq\lq popularity\rq\rq. Based on this simple mechanism, we study
two kinds of weighted networks, weighted small-world and weighted
scale-free, in which each link $ij$ in the network has an
associated weight $(k_{i}k_{j})^{\theta}$, where $k_{i}$ and
$k_{j}$ are the node degrees at the endpoint of the link $ij$. In
contrast to adding weights to links during networks being
constructed, we only consider weights depending on the \lq\lq
popularity\rq\rq of the nodes represented by their connectivity.
It was found that the both weighted networks have broad
distributions on characterization their link weights and vertex
strength, an exponential decay for the WWS network and a power law
behavior for the WBA network. The average shortest path length of
the WWS network remain invariable comparing with the WS network.
Whereas, the average shortest path length of the WBA network is
about $\langle k \rangle^{2\theta}$ times as that of the BA
network. Furthermore, as a survey of the model, we investigated
the spread of infectious diseases in both weighted networks based
on the standard SI model. The spreading velocity reaches a peak
very quickly after the infection outbreaks in both weighted
networks. Moreover, in contrast to trivial velocity decay for the
WWS network, an exponential decay was found for the WBA network.

As networks play an increasing role in the exploration of complex
systems, there is an imminent need to understand the interplay
between network dynamics and topology. We model the weights as
solely dependent on the topology, potentially overlooking
correlations among the weights themselves and the coupled
evolution in time. Recently, Barrat \emph{et al.} presented a
model \cite {Barrat_2} to study the dynamical evolution of weights
according to the topological variations, which is a good starting
point towards uncovering the role of such mechanisms.

\begin{acknowledgments}
This work was partly supported by the Natural Science Foundation
of Gansu Province (ZS011-A25-004-2) and the Doctoral Research
Foundation awarded by Lanzhou University.
\end{acknowledgments}

\newpage

\bigskip

\begin{references}

\bibitem{Strogatz}
S.H. Strogatz, Nature \textbf{410}, 268 (2001).

\bibitem{Barabasi_1}
A.-L. Barab\'{a}si and R. Albert, Rev. Mod. Phys. \textbf{74}, 47
(2002).

\bibitem{Dorogovtsev}
S.N. Dorogovtsev and J.F.F. Mendes, Advances in Physics
\textbf{51}, 1079 (2002).

\bibitem{Newman_0}
M.E.J. Newman, SIAM Review \textbf{45}, 167 (2003).

\bibitem{Cancho}
R.F.I. Cancho and R.V. Sol\'{e}, Proc. R. Soc. London, Ser. B
\textbf{268}, 2261 (2001).

\bibitem{Sigman}
M. Sigman and G.A. Cecchi, Proc. Natl. Acad. Sci. U.S.A.
\textbf{99}, 1742 (2002).

\bibitem{Newman_1}
M.E.J. Newman, Proc. Natl. Acad. Sci. U.S.A. \textbf{98}, 404
(2001); Phys. Rev. E \textbf{64}, 016131, 016132 (2001).

\bibitem{Barabasi_2}
A.-L. Barab\'{a}si, H. Jeong, Z. N\'{e}da, E. Ravasz, A. Schubert,
and T. Vicsek, Physica A \textbf{311}, 590 (2002).

\bibitem{Albert}
R. Albert, H. Jeong, and A.-L. Barab\'{a}si, Nature \textbf{401},
130 (1999).

\bibitem{Huberman_1}
B.A. Huberman and L.A. Adamic, Nature \textbf{401}, 131 (1999).

\bibitem{Huberman_2}
B.A. Huberman, P.L.T. Pirolli, J.E. Pitkow, and R.M. Lukose,
Science \textbf{280}, 95 (1999).

\bibitem{Caldarelli}
G. Caldarelli, R. Marchetti, and L. Pietronero, Europhys. Lett.
\textbf{52}, 386 (2000).

\bibitem{Amaral}
L.A.N. Amaral, A. Scala, M. Barthelemy, and H.E. Stanley, Proc.
Natl. Acad. Sci. U.S.A. \textbf{97}, 149 (2000).

\bibitem{McCann}
K. McCann, A. Hastings, and G.R. Huxel, Nature \textbf{395}, 794
(1998).

\bibitem{Williams}
R.J. Williams and N.D. Martinez, Nature \textbf{404}, 180 (2000).

\bibitem{Alon}
U. Alon, M.G. Surette, N. Barkai, and S. Leibler, Nature
\textbf{397}, 168 (1999).

\bibitem{Jeong_1}
H. Jeong, B. Tombor, R. Albert, Z.N. Oltvai, and A.-L.
Barab\'{a}si, Nature \textbf{407}, 651 (2000).

\bibitem{Jeong_2}
H. Jeong, S.P. Mason, A.-L. Barab\'{a}si, and Z.N. Oltvai, Nature
\textbf{411}, 41 (2001).

\bibitem{Watts_1}
D.J. Watts and S.H. Strogatz, Nature \textbf{393}, 440 (1998).

\bibitem{Barabasi_3}
A.-L. Barab\'{a}si and R. Albert, Science \textbf{286}, 509
(1999); A.-L. Barab\'{a}si, R. Albert, and H. Jeong, Physica A
\textbf{272}, 173 (1999).

\bibitem{Yook}
S.H. Yook, H. Jeong, and A.-L. Barab\'{a}si, and Y. Tu, Phys. Rev.
Lett. \textbf{86}, 5835, (2001).

\bibitem{Zheng}
D. Zhang, S. Trimper, B. Zheng, and P.M. Hui, Phys. Rev. E 67,
040102, (2003).

\bibitem{Barrat_1}
A. Barrat, M. Barth\'{e}lemy, R. Pastor-Satorras, and A.
Vespignai, Proc. Natl. Acad. Sci. U.S.A. \textbf{101}, 3747
(2004).

\bibitem{Park}
K. Park, Y.-C. Lai, and N. Ye, Phys. Rev. E \textbf{70}, 026109
(2004).

\bibitem{Goh_1}
K.-I. Goh, B. Kahng, and D. Kim, Phys. Rev. Lett. \textbf{87},
278701 (2001).

\bibitem{Szabo}
G. Szab\'{o}, M. Alava, and J. Kert\'{e}sz, Phys. Rev. E
\textbf{66}, 026101 (2002).

\bibitem{Kuperman}
M. Kuperman and G. Abramson, Phys. Rev. Lett. \textbf{86}, 2909
(2001).

\bibitem{Pastor}
R. Pastor-Satorras and A. Vespignani, Phys. Rev. Lett.
\textbf{86}, 3200 (2001); Phys. Rev. E \textbf{63}, 066117 (2001).

\bibitem{Newman_3}
M.E.J. Newman, Phys. Rev. E \textbf{66}, 016128 (2002).

\bibitem{Murray}
J.D. Murray, \textit{Mathematical Biology} (Springer Verlag,
Berlin, 1993).

\bibitem{Yan}
G. Yan, T. Zhou, J. Wang, Z.-Q. Fu, and B.-H. Wang, Chin. Phys.
Lett. \textbf{22}, 510 (2005).

\bibitem{Barrat_2}
A. Barrat, M. Barth\'{e}lemy, and A. Vespignani, Phys. Rev. Lett.
\textbf{92}, 228701 (2004).

\end{references}
\end{document}